\RequirePackage{fix-cm}

\documentclass[smallextended]{svjour3}       
\smartqed  
\usepackage{graphicx}
\usepackage{xcolor}

\newcommand{\be}{\begin{equation}}
\newcommand{\ee}{\end{equation}}
\newcommand{\bea}{\begin{eqnarray}} 
\newcommand{\eea}{\end{eqnarray}}

\usepackage{tensor}
\begin{document}

\title{Tracing the high energy theory of gravity: an introduction to Palatini inflation}


\author{Tommi Tenkanen
}


\institute{T. Tenkanen \at
              \,Department of Physics and Astronomy\\ Johns Hopkins University \\
3400 N. Charles Street\\ Baltimore, MD 21218\\ USA\\
              \email{ttenkan1@jhu.edu}        
}

\date{
}

\maketitle

\begin{abstract}
We present an introduction to cosmic inflation in the context of {\it Palatini} gravity, which is an interesting alternative to the usual {\it metric} theory of gravity. In the latter case only the metric $g_{\mu\nu}$ determines the geometry of space-time, whereas in the former case both the metric and the space-time connection $\Gamma^\lambda_{\mu\nu}$ are {\it a priori} independent variables -- a choice which can lead to a theory of gravity different from the metric one. In scenarios where the field(s) responsible for cosmic inflation are coupled non-minimally to gravity or the gravitational sector is otherwise extended, assumptions of the underlying gravitational degrees of freedom can have a big impact on the observational consequences of inflation. We demonstrate this explicitly by reviewing several interesting and well-motivated scenarios including Higgs inflation, $R^2$ inflation, and $\xi$-attractor models. We also discuss some prospects for future research and argue why $r=10^{-3}$ is a particularly important goal for future missions that search for signatures of primordial gravitational waves.
\end{abstract}


\section{Introduction}
\label{introduction}

In this paper we present a simple step-by-step introduction to cosmic inflation in the context of so-called Palatini gravity, review some scenarios previously studied in the literature, and discuss a particularly important goal for future missions that search for signatures of primordial gravitational waves. We assume the reader is familiar with the basics of General Relativity (GR) and knows at least some cosmology, in particular cosmic inflation, but otherwise no prior knowledge about different gravitational theories is required. 

While the space of all possible theories of gravity is vast and other, often equally well-motivated choices could be made, here we concentrate on the differences between simple "metric" and "Palatini" theories only, in particular in the context of cosmic inflation. We present a GR-based introduction to the topic, highlighting some subtle differences between the metric and Palatini theories of gravity in different scenarios, and then move on to discuss cosmic inflation. We pay particular attention to inflationary observables and their dependence on the assumptions of the underlying theory of gravity, demonstrating the differences between the metric and Palatini theories in several interesting scenarios including the famous Higgs inflation, $R^2$ inflation, and $\xi$-attractors, all of which are models that have been actively studied within the cosmology community during the past few years.

While the Palatini theories are almost as old as the usual metric theories of gravity\footnote{The "Palatini formulation" of GR is usually credited to the paper \cite{Palatini} by Attilio Palatini, but actually it was, apparently, first presented in the paper \cite{Einstein} by Albert Einstein \cite{Ferraris}. We will discuss the Palatini formulation of GR in more detail in the next section.} and the idea of cosmic inflation is already four decades old, inflation in the context of Palatini gravity has attained interest only relatively recently. The idea is often credited to a paper by Bauer and Demir \cite{Bauer:2008zj}, however see Refs. \cite{Guendelman:1998cz,Guendelman:2000rt,Kaganovich:2000fc,Meng:2003bk,Meng:2004yf,Allemandi:2004ca,Sotiriou:2005hu,Sotiriou:2005cd,Poplawski:2005sc} for some earlier work on the topic, mainly in the context of "quintessential inflation" where the same scalar field that drives the early-time inflation is also responsible for the late-time cosmic acceleration, but also in the context of $R^2$ inflation in Palatini theory \cite{Meng:2003bk,Meng:2004yf}. Recently, however, the topic has attained increasing interest within the community \cite{Bauer:2010jg,Tamanini:2010uq,Enqvist:2011qm,Borowiec:2011wd,Borowiec:2015qrp,Szydlowski:2015fcq,Stachowski:2016zio,Azri:2017uor,Rasanen:2017ivk,Fu:2017iqg,Tenkanen:2017jih,Racioppi:2017spw,Markkanen:2017tun,Jarv:2017azx,Racioppi:2018zoy,Azri:2018gsz,Enckell:2018kkc,Carrilho:2018ffi,Enckell:2018hmo,Antoniadis:2018ywb,Rasanen:2018fom,Kannike:2018zwn,Rasanen:2018ihz,Almeida:2018oid,Antoniadis:2018yfq,Takahashi:2018brt,Jinno:2018jei,Shimada:2018lnm,Tenkanen:2019jiq,Rubio:2019ypq,Jinno:2019und,Giovannini:2019mgk,Tenkanen:2019xzn,Benisty:2019tno,Bostan:2019uvv,Bostan:2019wsd,Tenkanen:2019wsd,Gialamas:2019nly,Racioppi:2019jsp,Bostan:2019yep,Shaposhnikov:2020geh,Tenkanen:2020cvw} and, as we will discuss, there are still many interesting aspects that remain to be studied.

The paper is organized as follows: we begin in Sec. \ref{geometry} by a short review of constructing the space-time geometry, mainly following the notation and conventions of Ref. \cite{Carroll:2004st}, and then move on to discuss cosmic inflation in Sec. \ref{inflation}. We will first present a general overview of the basics of inflation, then discuss some of the most important observables in Sec. \ref{observations}, and then study some classes of models in more detail in Sec. \ref{models}. In Sec. \ref{future_missions}, we will discuss a particularly important goal for future missions that search for signatures of primordial gravitational waves. In Sec. \ref{summary}, we briefly summarize the discussion and ponder some directions for future research.


\section{Constructing the space-time geometry}
\label{geometry}

In curved space-time, partial derivatives are not sufficient to describe how objects such as vectors change from point to point, as they depend on the coordinate system used. One would therefore like to define a similar operator but in a way independent of coordinates. By requiring that such an operator, called {\it covariant derivative} and denoted by $\nabla_\mu$, is linear and follows the usual product rule, one finds the covariant derivative of an arbitrary vector $V^\mu$ to be given by
\begin{equation}
\label{covariant_derivative}
\nabla_\mu V^\nu = \partial_\mu V^\nu + \Gamma^\nu_{\mu\lambda}V^\lambda\,,
\end{equation}
where $ \Gamma^\nu_{\mu\lambda}$ is called a {\it connection}, and which ensures that the definition of $\nabla_\mu$ is independent of coordinates. The definition (\ref{covariant_derivative}) generalizes to tensors of arbitrary rank 
\begin{eqnarray}
\label{covariant_derivative_general}
\nabla_\sigma T\indices{^{\mu_1\mu_2\dots \mu_k}_{\nu_1\nu_2\dots\nu_l}} &=& \partial_\sigma  T\indices{^{\mu_1\mu_2\dots \mu_k}_{\nu_1\nu_2\dots\nu_l}} \\ \nonumber
&+& \Gamma^{\mu_1}_{\sigma\lambda} T\indices{^{\lambda\mu_2\dots \mu_k}_{\nu_1\nu_2\dots\nu_l}}
+ \Gamma^{\mu_2}_{\sigma\lambda} T\indices{^{\mu_1\lambda\dots \mu_k}_{\nu_1\nu_2\dots\nu_l}} 
+ \dots \\ \nonumber
&-& \Gamma^{\lambda}_{\sigma\nu_1} T\indices{^{\mu_1\mu_2\dots \mu_k}_{\lambda\nu_2\dots\nu_l}}
 - \Gamma^{\lambda}_{\sigma\nu_2} T\indices{^{\mu_1\mu_2\dots \mu_k}_{\nu_1\lambda\dots\nu_l}}
 - \dots \,,
\end{eqnarray}
when one also requires that the covariant derivative commutes with contractions and reduces to partial derivatives on scalars. In particular, this definition applies to the space-time {\it metric} $g_{\mu\nu}$ (by {\it space-time} we refer to a 4-dimensional manifold equipped with a connection $ \Gamma^\sigma_{\mu\nu}$ and a symmetric metric tensor $g_{\mu\nu}=g_{\nu\mu}$). Note, however, that the connection itself is not a tensor, as it does not transform in coordinate transformations as proper tensors (such as the metric) do. This is intentional, as the connection has been constructed in such a way that the covariant derivative (\ref{covariant_derivative_general}) transforms as tensors do.

While the above requirements are enough to construct a derivative operator independent of coordinates, in the context of GR one usually postulates that the connection is also
\begin{enumerate}
\item {\it torsion-free}, $\Gamma^\lambda_{\mu\nu} = \Gamma^\lambda_{\nu\mu}$\,, and
\item {\it metric-compatible}, $\nabla_\rho g_{\mu\nu} = 0$\,.
\end{enumerate}
It is easy to show that together these two postulates determine the space-time connection uniquely in terms of the metric:
\be
\label{Levi-Civita}
	\bar{\Gamma}^\sigma_{\mu\nu} = \frac{1}{2}g^{\sigma\rho}(\partial_\mu g_{\nu\rho} + \partial_\nu g_{\rho\mu} - \partial_\rho g_{\mu\nu}) \,.
\ee
This is the famous Levi-Civita connection encountered in GR\footnote{For clarity, we note that sometimes the notation $\{^\sigma_{\mu\nu}\}$ is used instead of $\bar{\Gamma}^\sigma_{\mu\nu}$, and the name "Christoffel symbol" or "Riemannian connection" instead of the "Levi-Civita connection" we use in this paper.}. However, as the simple definition of the covariant derivative suggests, no metric is necessary to construct some aspects of geometry on a space-time -- that is, the notions of metric and connection are not necessarily intertwined.  Therefore, we denote the special Levi-Civita connection with an overbar to distinguish it from other connections that do not satisfy the two postulates above, as it is easy to think of other gravitational theories which do not satisfy them {\it a priori}. We will give examples of such theories below.

Before doing so, let us introduce one more concept. In curved space, the result of parallel transporting\footnote{Parallel transporting a tensor $T$ along the path $x^\mu(\lambda)$ parameterized by $\lambda$ means that the covariant derivative of $T$ along the path vanishes,
\begin{equation}
\frac{{\rm d}x^\sigma}{{\rm d}\lambda}\nabla_\sigma T\indices{^{\mu_1\mu_2\dots \mu_k}_{\nu_1\nu_2\dots\nu_l}} = 0\,.
\end{equation}
Also this quantity depends, {\it a priori}, only on the connection $\Gamma^\sigma_{\mu\nu}$. Note, however, that if the connection is not metric-compatible, parallel transport does not necessarily preserve the norm of vectors, a feature often taken as granted.
} a tensor of arbitrary rank from one point to another will depend on the path taken between the points. For simplicity, consider a vector $V^\rho$ parallel transported along a loop: the corresponding change is described by
\begin{equation}
\delta V^\rho = R\indices{^\rho_{\sigma\mu\nu}} V^\sigma A^\mu B^\nu\,,
\end{equation}
where $A^\mu$ and $B^\nu$ are vectors that define the loop and
\be
R\indices{^\rho_{\sigma\mu\nu}} = \partial_\mu\Gamma^\rho_{\nu\sigma} - \partial_\nu\Gamma^\rho_{\mu\sigma} + \Gamma^\rho_{\mu\lambda}\Gamma^\lambda_{\nu\sigma}  - \Gamma^\rho_{\nu\lambda}\Gamma^\lambda_{\mu\sigma} \,,
\ee
is the {\it Riemann tensor} which describes the change experienced by tensors of arbitrary rank when parallel transported due to curvature of the space-time. This definition holds for {\it any} connection, whether or not it is metric-compatible or torsion-free. However, for an arbitrary connection the Riemann tensor has only one obvious symmetry: it is antisymmetric in the last two indices. The rest of the standard symmetries are not present for an arbitrary connection \cite{Sotiriou:2006qn}. Most importantly for the purposes of this paper, however, we note again that no metric is needed to define the Riemann tensor.

Let us then see how this works in the case of GR. As is well known, the Lagrangian formulation of GR is encoded in the simple action
\be \label{GR_action}
	S = \frac{1}{2} M_{\rm P}^2\int d^4x \sqrt{-g}g^{\mu\nu}R_{\mu\nu}(\Gamma)\,,
\ee
which is often called the Einstein-Hilbert action. It represents the minimal choice of degrees of freedom -- up to a boundary term\footnote{Derivation of the equations of motion for the metric case requires adding a so-called Gibbons-Hawking-York boundary term to the action to cancel a total derivative term that depends on the second derivatives of the metric \cite{York:1972sj,Gibbons:1976ue}. For the possibility of adding a (non-covariant) boundary term that only depends on the first derivatives of the metric, see e.g. Ref. \cite{Dyer:2008hb}.} -- and ensures that the resulting field equations are second order differential equations. Here $M_{\rm P} = 1/\sqrt{8\pi G}$ is the reduced Planck mass in natural units and $G$ is the Newton's gravitational constant, $g$ is the determinant of the metric tensor, the {\it Ricci tensor} is constructed from the Riemann tensor by contraction\footnote{While this definition of the Ricci tensor is not unique, the definition of the curvature scalar (\ref{Ricci_scalar}) is \cite{Sotiriou:2006qn}.}, $R_{\mu\nu} = R\indices{^{\lambda}_{\mu\lambda\nu}}$, and $\Gamma$ is a shorthand notation for the three-index connection. Usually the action is written in terms of the curvature (Ricci) scalar
\begin{equation}
\label{Ricci_scalar}
R \equiv g^{\mu\nu}R_{\mu\nu}(\Gamma)\,,
\end{equation}
which often contains the implicit assumption that the connection is metric-compatible. If this was the case, we say that the theory is of {\it metric} type. In a so-called {\it Palatini} theory, however, both $g_{\mu\nu}$ and $\Gamma$ are treated as independent variables without the assumptions of metric compatibility or the usual symmetries for the Riemann tensor, and the metric appears as an auxiliary variable with no kinetic term. In this case, then, variation of the action with respect to the metric gives the usual Einstein's equations but for a Ricci tensor constructed from a connection that has no {\it a priori} relationship with the metric. It is an interesting exercise to show that when the connection is torsion-free, variation of the action with respect to the connection leads to the requirement that it must also be metric-compatible, i.e. the Levi-Civita connection (\ref{Levi-Civita}), and therefore in this case the Einstein's equations will be exactly equal to those in the metric case. Therefore, for the Einstein-Hilbert action (\ref{GR_action}) the Palatini and metric theories render to mere formulations of the {\it same theory}, i.e. the GR. However, as discussed in e.g. Refs. \cite{Rasanen:2017ivk,Demir:2020brg}, it could be argued that the Palatini formulation is much simpler than the metric formulation, because there is no need to add a boundary term to the action, as it involves only first derivatives of the variables that are to be varied over. For this reason, the Palatini formalism is also sometimes called the "first order formalism", whereas the metric formulation is dubbed as the "second order formalism".

However, with non-minimally coupled matter fields or otherwise enlarged gravity sector, the two approaches do not generally correspond to different formulations of the same theory. For example, in a theory which contains a {\it non-minimally coupled} scalar field $\phi$, specified by the action
\be \label{nonminimal_action1}
	S = \int d^4x \sqrt{-g}\left(\frac{1}{2}\left(M_{\rm P}^2 + \xi \phi^2\right) g^{\mu\nu}R_{\mu\nu}(\Gamma) - \frac{1}{2} g^{\mu\nu}\nabla_{\mu}\phi\nabla_{\nu}\phi - V(\phi) \right) \,,
\ee
where $\xi$ is a dimensionless coupling constant, $\nabla_\mu$ is the covariant derivative, $V(\phi)$ is the potential for the scalar field, and $\Gamma$ is assumed to be, for simplicity, torsion-free, variation of the action with respect to the connection gives
\be
\label{connection_solution}
\Gamma^\sigma_{\mu\nu} = \bar{\Gamma}^\sigma_{\mu\nu} + \delta^\sigma_\mu\partial_\nu\omega(\phi) + \delta^\sigma_\nu\partial_\mu\omega(\phi) - g_{\mu\nu}\partial^\sigma\omega(\phi) ,
\ee
where 
\begin{equation}
\label{w}
\omega(\phi)=\ln\sqrt{1+\xi\frac{\phi^2}{M_{\rm P}^2}}\,. 
\end{equation}
Because the result (\ref{connection_solution}) clearly differs from the Levi-Civita connection (\ref{Levi-Civita}) for any finite value of $\phi$, we conclude that in the case of non-minimally coupled scalar fields the Palatini and metric theories do not represent different formulations of the same theory but are two entirely different theories. This is true for otherwise enlarged gravity sectors as well, such as those including terms higher order in $R$ \cite{Sotiriou:2006qn,Sotiriou:2008rp}. Therefore, whenever a model contains non-minimal gravitational couplings, one has to make a {\it choice} regarding the underlying gravitational degrees of freedom. Only observations can tell which option is the one actually realized in Nature.

However, two things are worth making a remark on. First, while the metric approach could seem simpler and is the one usually considered in the literature (despite the need for the extra boundary term), at least in the context of cosmic inflation, choosing the Palatini approach does not necessarily amount to adding new (dynamical) degrees of freedom to the theory. Moreover, this choice can actually simplify calculations, as we will see. Second, as Eq. (\ref{w}) shows, when the scalar field relaxes to zero, $\phi \to 0$, as is the case in most inflationary models for the dynamics after inflation, one retains the pure GR form of the theory regardless of the original choice between the metric and Palatini theory. In practice, however, even a milder requirement is sufficient: one can allow $\phi \to v$, where $v\ll M_{\rm P}$ or even just $\Delta\phi/M_{\rm P}\ll 1$ (at late times), without inducing any observable deviation from GR \cite{Dvali:2001dd}. In this way, one can modify the way gravity works in the early universe without modifying it at late times.

Due to these differences between the metric and Palatini gravity, it is an interesting starting point to consider a theory where geometry, in particular the space-time connection, depends on both the metric and the matter fields coupled non-minimally to gravity, especially in the context of cosmic inflation. As we will show, models of inflation in the Palatini and metric formulations are intrinsically different, which has profound consequences on inflationary observables and thus also on our ability to test the underlying gravitational degrees of freedom at high energies. 


\section{Cosmic inflation}
\label{inflation}

Let us consider the action 
\begin{equation}
\label{jordanframe}
S_J = \int d^4x \sqrt{-g}\left(\frac12M_{\rm P}^2\Omega^2(\phi) g^{\mu\nu}R_{\mu\nu}(\Gamma) - \frac{1}{2}K(\phi) g^{\mu\nu}\nabla_{\mu}\phi\nabla_{\nu}\phi - V(\phi) \right)\,,
\end{equation}
where $\Omega^2(\phi)$ and $K(\phi)$ are non-singular functions of the scalar field $\phi$, and which is a slight generalization of the action (\ref{nonminimal_action1}) discussed in the previous section. However, here we assume, for simplicity, that the gravity part of the action is still given by
\begin{equation}
\label{Omega_F}
\Omega^2(\phi) = 1 + \frac{F(\phi)}{M_{\rm P}^2}\,,
\end{equation} 
where $F(\phi)$ is again a non-singular function of the scalar field and the first term represents the usual Einstein-Hilbert choice\footnote{Scenarios where this term is absent and the scalar field is responsible for generating the Einstein-Hilbert term are sometimes called "induced gravity" models \cite{Spokoiny:1984bd,Accetta:1985du,Kaiser:1993bq,Kaiser:1994wj}.}. While this is the choice of $\Omega(\phi)$ most often considered in the literature, other interesting scenarios have been studied in the literature too; see e.g. Refs. \cite{Enckell:2018hmo,Antoniadis:2018ywb,Antoniadis:2018yfq,Tenkanen:2019jiq,Tenkanen:2019wsd,Gialamas:2019nly} for the inclusion of an $R^2$ term and Refs. \cite{Rasanen:2018ihz,Shimada:2018lnm} for non-vanishing torsion and kinetic terms for the metric in the Palatini case, which introduce multiple extra terms to the action. In Sec. \ref{models}, we will briefly discuss the former case above.

Because in the frame (\ref{jordanframe}) the non-minimal coupling between the scalar field and gravity is explicit, this frame is called the Jordan frame; hence the subscript in $S_J$. Note that when $F \to 0$, we retain the GR regardless of the choice of the approach, i.e. whether we want to consider metric or Palatini gravity. However, at $F\neq 0$, the two approaches correspond to different theories. To ease the comparison with the usual metric approach, in the following we will present the results for both the metric and Palatini cases. We will also omit the arguments, $\Omega \equiv \Omega(\phi)$, $K\equiv K(\phi)$, and $F\equiv F(\phi)$, for simplicity. 

While we could proceed by simply varying the action (\ref{jordanframe}) with respect to our degrees of freedom to derive the evolution or constraint equations for them, it is often much simpler to analyze inflation in a frame where the gravity sector is canonical, i.e. where effectively $\Omega=1$ and $\Gamma=\bar{\Gamma}$. This can be achieved by performing a Weyl transformation
\be \label{Omega1}
	\bar{g}_{\mu\nu} \equiv \Omega^2 g_{\mu\nu}\,,
\ee
which makes the relevant quantities transform as
\begin{eqnarray}
\sqrt{-g} &=& \Omega^{-4}\sqrt{-\bar{g}} \\ 
R&=&  \Omega^2\left(1 - \kappa \times 6\Omega\bar{\nabla}^\mu\bar{\nabla}_\mu\Omega^{-1} \right)\bar{R}\,,
\end{eqnarray}
where $\kappa=1$ in the metric case and $\kappa=0$ in the Palatini case, and the quantities with an overbar are defined in terms of the new metric $\bar{g}_{\mu\nu}$. Then, for the choice of $\Omega$ as in Eq. (\ref{Omega_F}) and after partial integration and some algebra, the action (\ref{jordanframe}) becomes 
\be \label{einsteinframe1}
	S_E = \int d^4x \sqrt{-\bar{g}}\left(\frac{1}{2}M_{\rm P}^2 \bar{R} - \frac12\left(\frac{K}{\Omega^2} + \kappa\times \frac32\frac{F_{,\phi}^2}{\Omega^4 M^2_{\rm P}} \right)\bar{\nabla}_{\mu}\phi\bar{\nabla}^\mu\phi - \frac{V(\phi)}{\Omega^4} \right) \,,
\ee
where the subscript of $F$ denotes derivative with respect to the field $\phi$. Because in this frame the non-minimal coupling to gravity vanishes, this frame is called the Einstein frame; hence the subscript in $S_E$. Because now the connection appears only in the usual Einstein-Hilbert term, in this frame $\Gamma = \bar{\Gamma}$, i.e. one retains the Levi-Civita connection, and therefore $\bar{R} \equiv R(\bar{\Gamma})$ and $\bar{g}_{\mu\nu}$ is the metric compatible with the connection $\bar{\Gamma}$. Now the gravity sector is canonical.

If only one scalar field is dynamical, also the scalar field kinetic term in (\ref{einsteinframe1}) can be expressed in a canonical form with a suitable field redefinition $\phi = \phi(\chi)$, determined by 
\be \label{chi1}
	\frac{d\phi}{d\chi} = \sqrt{\frac{\Omega^4}{K\Omega^2 +\kappa\times \frac32 F_{,\phi}^2/M^2_{\rm P}}} \,.
	\ee
For multifield scenarios, see Refs. \cite{Lerner:2009xg,Kaiser:2013sna,Tenkanen:2016idg,Carrilho:2018ffi,Almeida:2018oid,Tenkanen:2019xzn}. The action~(\ref{einsteinframe1}) then becomes
\be \label{EframeS1}
	S_{\rm E} = \int d^4x \sqrt{-g}\bigg(\frac{1}{2}M_{\rm P}^2\bar{R} -\frac{1}{2}{\bar{\nabla}}_{\mu}\chi{\bar{\nabla}}^{\mu}\chi - U(\chi)  \bigg) \,,
\ee
where 
\begin{equation}
U(\chi) = \frac{V(\phi(\chi))}{\Omega^{4}(\phi(\chi))}\,,
\end{equation}
where the arguments emphasize that the potential is for the field $\chi$, even if one was not able to find an analytical expression for it in terms of $\phi$. 

Now, if the scalar field's potential energy dominated the total energy density in the early universe and the potential is flat enough, for example if it develops a plateau at large field values, it will be suitable for inflation and may allow to generate the observed spectrum for curvature perturbations and thus seed the origins of large scale structure in the universe. If this was the case, we call the scalar field an {\it inflaton} field. However, before discussing the inflationary dynamics and the resulting observables in more detail, let us again make a few remarks regarding the action (\ref{EframeS1}).

First, the result (\ref{EframeS1}) clearly exhibits the fact that single-field Palatini models with vanishing torsion are equivalent to metric theories with a different choice of the potential for the scalar field in the Einstein frame. Indeed, the only difference between the two theories of gravity is in this case in the value of $\kappa$, which affects the re-definition of the field (\ref{chi1}) and thus also the potential for the canonically normalized field $\chi$. Had we started in the Einstein frame and bluntly put a given potential in by hand, the result would have been exactly the same as in the case where we started in the Jordan frame with a suitable potential and performed the Weyl transformation and field re-definition. This reflects the fact that this kind of Palatini theories are {\it metric-affine}, i.e. nothing but metric theories in disguise \cite{Koivisto:2005yc,Sotiriou:2006qn,Capozziello:2007ec,Capozziello:2015lza,Azri:2017uor,Aoki:2018lwx,Azri:2018gsz,Rasanen:2018ihz,Shimada:2018lnm}. 

However, one can rightfully argue that certain choices are better motivated than others. First, as usual quantum field theory (QFT) reveals, only potentials up to mass dimension four are renormalizable in flat space. Second, non-minimal couplings to gravity should be seen not as an {\it ad hoc} addition to scalar field models but as a generic ingredient of QFT in a curved space-time, both because they are generated radiatively and because no symmetry arguments forbid including them in the theory in the first place \cite{Birrell:1982ix}. It is in this sense that one can say that the differences between the scenarios we call "metric" or "Palatini" are indeed in the underlying theory of gravity and not in the choice of the scalar field kinetic term or potential. While at classical level one could, in principle, start in the Einstein frame with any scalar potential or kinetic term, it can be argued that it is the surprising connection with gravity that makes only certain models particularly interesting and the underlying gravitational degrees of freedom testable, as we will discuss in more detail in the next sections.


\subsection{Inflationary dynamics and observables}
\label{observations}

Let us then consider the dynamics during inflation and, most importantly, observables. We assume that in the early universe the scalar field $\chi$ was energetically dominant, so that in {\it slow-roll} approximation where the field moves slowly down in its potential, $|\ddot{\chi}|\ll 3H|\dot{\chi}|\,, \dot{\chi}^2 \ll U$, where $H\equiv \dot{a}/a$ is the Hubble parameter and the dot denotes derivative with respect to time, the Friedmann equation and the inflaton equation of motion (found by varying the action (\ref{EframeS1}) with respect to the metric and the field $\chi$, respectively) become
\begin{eqnarray}
H^2 &\simeq& \frac{U}{3M_{\rm P}^2}\,, \\ 
3H\dot{\chi} &\simeq& -U_{,\chi}\,,
\end{eqnarray}
where the subscript of $U$ denotes derivative with respect to $\chi$. The inflationary dynamics can then be characterized by the usual slow-roll parameters
\be
\label{SRparameters1}
	\epsilon \equiv \frac{1}{2}M_{\rm P}^2 \left(\frac{U'}{U}\right)^2 \,, \quad
	\eta \equiv M_{\rm P}^2 \frac{U''}{U} \,,
\ee
so that in slow-roll $\epsilon\,, |\eta | \ll 1$. Then also $\epsilon = -\dot{H}/H^2 \ll 1$, and we have $H\sim$ constant, $a\sim e^{Ht}$, and thus also $\ddot{a}>0$ -- the expansion of the universe is accelerating (the universe is "inflating"). For discussion on inflationary dynamics beyond the slow-roll paradigm, see e.g. Ref.~\cite{Lyth:1998xn}.

Another important and useful quantity is the number of $e$-folds between the horizon exit of the scale where measurements are made (the "pivot" scale) and the end of inflation, $N$, which is given by
\be \label{Ndef}
	N = \frac{1}{M_{\rm P}^2} \int_{\chi_{\rm end}}^{\chi_*} {\rm d}\chi \, U \left(\frac{{\rm d}U}{{\rm d} \chi}\right)^{-1}.
\ee
The field value at the end of inflation, $\chi_{\rm end}$, is defined via $\epsilon(\chi_{\rm end})= 1$, as this criterion signals the end of slow-roll and accelerating expansion. The field value at the time when the pivot scale exited the horizon is denoted by $\chi_*$. In the following, we will use $k_*=0.05\, {\rm Mpc}^{-1}$ as the pivot scale.

To explain the famous horizon and flatness problems (see e.g. Ref. \cite{Baumann:2009ds}), inflation should have lasted for a sufficiently long time, $N\gg 1$. The exact number depends on the post-inflationary expansion history, especially on the details of (p)reheating that follow inflation and thermalize the universe; see Refs. \cite{Allahverdi:2010xz,Amin:2014eta} for a review. While usually $N\sim 50 - 60$ is assumed, the number can be as low as $N\sim 20$ \cite{Liddle:2003as}. However, it is not enough for an inflationary model to provide for a sufficient number of e-folds, as the inflaton field also acquires fluctuations and thus seeds the origins of structure at different scales\footnote{Unless the majority of metric perturbations is generated in some other way, for example via the curvaton \cite{Enqvist:2001zp,Lyth:2001nq,Moroi:2001ct} or modulated reheating mechanism \cite{Dvali:2003em,Kofman:2003nx}. Here we do not consider such possibilities.}. At the largest physical distance scales, the scalar curvature power spectrum is observed to have a power-law form~\cite{Aghanim:2018eyx,Akrami:2018odb}
\be
\mathcal{P}_{\zeta}(k) = \mathcal{A}\left(\frac{k}{k_*}\right)^{n_s-1}\,,
\ee
where $\zeta$ denotes the curvature perturbation. The spectrum has the observed amplitude $\mathcal{A}\simeq 2.1\times 10^{-9}$ and {\it spectral tilt} (or {\it spectral index}) $n_s\simeq 0.965$ at the pivot scale $k_*=0.05\, {\rm Mpc}^{-1}$. 

Assuming slow-roll, the amplitude can be expressed as~\cite{Lyth:1998xn}
\be
\label{cobe}
 \mathcal{A} = \frac{1}{24 \pi^2 M_{\rm P}^4} \frac{U(\chi_*)}{\epsilon(\chi_*)} ,
\ee
and the leading order expression for the spectral tilt is
\bea
\label{nsralpha}
n_s -1\equiv \frac{{\rm d\,ln}\mathcal{P}_{\zeta}(k)}{{\rm d\,ln}k} \simeq -6\epsilon + 2\eta\,,
\eea
which can be used to relate the number of required $e$-folds to the potential and the model parameters, in particular in our case to the non-minimal coupling function $\Omega$, as we will show in the next subsections.

In addition to scalar curvature perturbations, the fluctuations of the inflaton field also generate tensor perturbations, i.e. gravitational waves. Their power spectrum is given by~\cite{Baumann:2009ds}
\begin{equation}
\mathcal{P}_{T} =  \frac{8}{M_{\rm pl}^2} \left( \frac{H}{2\pi}\right)^2\,,
\end{equation}
and the observational constraints are usually expressed in terms of the {\it tensor-to-scalar ratio}
\begin{equation}
\label{r_16e}
r \equiv \frac{\mathcal{P}_{T}}{\mathcal{P}_{\zeta}} \simeq 16\epsilon\,,
\end{equation}
where the last expression applies at the leading order in slow-roll parameters. As primordial tensor perturbations have not been discovered, observations of the Cosmic Microwave Background radiation (CMB) place an upper limit on tensor-to-scalar ratio, $r<0.06$ at the pivot scale $k_*=0.05\, {\rm Mpc}^{-1}$ \cite{Ade:2018gkx}. Again, this can be used to relate the number of required $e$-folds to the underlying model parameters including the non-minimal coupling function.

Finally, one can ask whether the spectral index or the tensor-to-scalar ratio remain constant or if they change over the observable range of scales. The change in $n_s$ is characterized by the running, $\alpha_s \equiv {\rm d}n_s/{\rm d\,ln}k$, and running of the running of the spectral index, $\beta_s \equiv {\rm d}^2n_s/{\rm d(ln}k)^2$, and the change in $r$ by the spectral index of gravitational waves, $n_T \equiv {\rm d\,ln}\mathcal{P}_{T}/{\rm d\,ln}k$. One can derive simple expressions also for these quantities in terms of slow-roll parameters similar to Eqs. (\ref{nsralpha}) and (\ref{r_16e}) but here we neglect them for simplicity, as the data are consistent with negligible $\alpha_s\,, \beta_s\,, n_T$~\cite{Akrami:2018odb}.


\subsection{Inflationary models}
\label{models}

So far, our discussion has been general in a sense that we have been agnostic of the inflaton potential. Let us now consider a few example cases to see how and to what extent observations can distinguish between different models and, in the context of inflationary models, different theories of gravity.

\subsubsection{Higgs inflation}

A particularly interesting scenario is the one where the Standard Model (SM) Higgs field is responsible for driving inflation. The SM Higgs is, after all, the only scalar field we know exists in Nature, and therefore attempting to realize inflation within the SM is arguably the simplest possibility. This scenario was originally considered in Ref. \cite{Bezrukov:2007ep} in the metric case (see also Refs. \cite{Spokoiny:1984bd,Futamase:1987ua,Salopek:1988qh,Fakir:1990eg,Amendola:1990nn,Kaiser:1994vs,CervantesCota:1995tz,Komatsu:1999mt} for earlier work on the topic) and in Ref. \cite{Bauer:2008zj} in the Palatini case. It has ever since attained considerable interest within the community; see Ref. \cite{Rubio:2018ogq} for a recent review on metric Higgs inflation and Refs. \cite{Bauer:2010jg,Rasanen:2017ivk,Enckell:2018kkc,Enckell:2018hmo,Rasanen:2018ihz,Takahashi:2018brt,Tenkanen:2019jiq,Rubio:2019ypq,Tenkanen:2019xzn,Bostan:2019wsd,Shaposhnikov:2020geh} for recent studies on Palatini-Higgs inflation.

The model is specified by
\begin{equation}
\label{Higgs_inflation_model}
\Omega^2 = 1 + \xi \frac{2(\Phi^\dagger\Phi)}{M_{\rm P}^2}\,, \quad
V = \lambda (\Phi^\dagger\Phi)^2\,,
\end{equation}
where $\sqrt{2}\Phi = (0,v+\phi)$ is the SM Higgs doublet in the unitary gauge, $v=246$ GeV is the electroweak (EW) scale, and $\xi$ is a dimensionless coupling constant. The choice (\ref{Higgs_inflation_model}) is particularly well-motivated, as even if the non-minimal coupling is not present at the classical level, it will be generated by quantum corrections \cite{Birrell:1982ix}. In the following, we take $V\approx \lambda/4 \phi^4$ where $\phi$ is the radial mode which corresponds to the physical Higgs field\footnote{For the Goldstone bosons, see Refs. \cite{Hertzberg:2010dc,Mooij:2011fi,Greenwood:2012aj,George:2013iia,George:2015nza}.} and which is assumed to be much larger than the EW scale during inflation, $\phi \gg v$.

By performing a Weyl transformation, we arrive at Eq. (\ref{einsteinframe1}), and can again re-define the field according to Eq. (\ref{chi1}). The solution to the resulting differential equation is \cite{GarciaBellido:2008ab,Bauer:2008zj,Rasanen:2017ivk}
\be
\label{chi_solution}
\frac{\sqrt{\xi}}{M_{\rm P}}\chi = \sqrt{1+6\kappa\xi}\sinh^{-1}\left(\sqrt{1+6\kappa\xi}u\right) - \sqrt{6\xi}\kappa\sinh^{-1}\left(\sqrt{6\xi}\frac{u}{\sqrt{1+u^2}}\right) ,
\ee
where $u\equiv \sqrt{\xi}\phi/M_{\rm P}$. Thus, at large field values
\bea
\phi(\chi) 
&\simeq& \displaystyle\frac{M_{\rm P}}{\sqrt{\xi}} \exp\left(\sqrt{\frac{1}{6}}\frac{\chi}{M_{\rm P}} \right)	  \quad \mathrm{metric} ,\\   
\phi(\chi) &=& \displaystyle\frac{M_{\rm P}}{\sqrt{\xi}}\sinh\left(\frac{\sqrt{\xi}\chi}{M_{\rm P}}\right) \quad \mathrm{Palatini} , 
\eea
so that the Einstein frame potential for the canonically normalized field becomes
\bea \label{chipotential1}
	U(\chi) 
&\simeq& \displaystyle\frac{\lambda M_{\rm P}^4}{4\xi^2}
\bigg(1+\exp\left(-\sqrt{\frac{2}{3}} \displaystyle\frac{\chi}{M_{\rm P}} \right) \bigg)^{-2}
	 \quad \mathrm{metric} ,\\ 
	U(\chi) &=& \displaystyle\frac{\lambda M_{\rm P}^4}{4\xi^2}
	\tanh^4\left(\displaystyle\frac{\sqrt{\xi}\chi}{M_{\rm P}}\right) \quad \mathrm{Palatini} , 
	\label{chipotential1_P}
\eea
where the expressions in the metric case apply for $\xi\gg 1$ and $\chi\gg \sqrt{3/2}M_{\rm P}$, and the expressions in the Palatini case are exact. We see that for $\chi\gg \sqrt{3/2}M_{\rm P}$ in the metric case or $\chi\gg M_{\rm P}/\sqrt{\xi}$ in the Palatini case the potential tends to a constant exponentially fast and is therefore suitable for slow-roll inflation.

It is now straightforward to compute the amplitude of the curvature perturbation power spectrum (\ref{cobe}) and its tilt (\ref{nsralpha}), as well as the predicted tensor-to-scalar ratio (\ref{r_16e}). For the amplitude one finds
\bea \label{xicondition1_M}
	\mathcal{A} &\simeq& 
	\displaystyle\frac{\lambda N^2}{72\pi^2\xi^2} \quad \mathrm{metric} , \vspace{3mm} \\
			\mathcal{A} &\simeq& \displaystyle\frac{\lambda N^2}{12\pi^2 \xi} \quad \mathrm{Palatini} ,
			\label{xicondition1_P}
\eea
whereas the spectral tilt and tensor-to-scalar ratio read
\begin{eqnarray}
\label{nsr_higgs_M}
n_s - 1&=& 
-\frac{2}{N} + \frac{3}{2N^2}\,  \quad \mathrm{metric} ,\\ 
n_s - 1&=& -\frac{2}{N} - \frac{3}{8\xi N^2}\, \quad \mathrm{Palatini} ,
\label{nsr_higgs_P}
\end{eqnarray}
\begin{eqnarray}
r &=& 
 \label{Higgs_nsr_M}
\frac{12}{N^2}-\frac{18}{N^3}\, \quad \mathrm{metric} ,\\ 
r &=& \frac{2}{\xi N^2}+\frac{1}{4\xi^2 N^3}\, \quad \mathrm{Palatini},
 \label{Higgs_nsr}
\end{eqnarray}
to second order in the expansion in $1/N$, where $N$ is the number of e-folds given by Eq. (\ref{Ndef}).

From Eqs. (\ref{nsr_higgs_M}), (\ref{nsr_higgs_P}) one sees that the predictions for the spectral index are essentially the same in the two theories. For example, for $N=60$ we obtain in both cases $n_s\simeq 0.967$, in agreement with the data. However, as Eqs. (\ref{xicondition1_M}), (\ref{xicondition1_P}) show, the relation $\mathcal{A}\propto \lambda/\xi^2$ encountered in the metric case gets modified to $\mathcal{A}\propto \lambda/\xi$ for the Palatini case. Therefore, not only the relation between the couplings $\lambda$ and $\xi$ changes from one gravitational theory to other but also one can use Eq. (\ref{Higgs_nsr}) to show that in the Palatini case the tensor-to-scalar ratio becomes $r\sim \mathcal{A}/(\lambda N^4)$, which can be orders of magnitude smaller than the corresponding quantity in the metric case, as Eq. (\ref{Higgs_nsr_M}) reveals. In the metric case $r\sim 10^{-3}$ for the usual number of e-folds\footnote{In principle, because the SM field content and couplings are known, the details of (p)reheating can be calculated exactly, which then gives the total number of e-folds between the end of inflation and horizon exit of the scale where measurements are made \cite{Bezrukov:2008ut,GarciaBellido:2008ab,Takahashi:2018brt,Rubio:2019ypq}. However, in practice the SM couplings are not exactly known, and neither is the Beyond-the-Standard-Model (BSM) physics which accommodates e.g. dark matter or baryogenesis, and which may affect the renormalization group running of the SM couplings up to the scales where (p)reheating (or inflation) occurs.}, $N\sim 50 - 60$, whereas in the Palatini case $r \sim 10^{-13}\dots 10^{-4}$, as studied in Refs. \cite{Rasanen:2017ivk,Takahashi:2018brt} (see also Ref. \cite{Shaposhnikov:2020geh}). Both predictions are consistent with the data. As further shown in  Refs. \cite{Rasanen:2017ivk,Takahashi:2018brt}, also the running, ${\rm d}n_s/{\rm d\,ln}k$, and running of the running of the spectral index, ${\rm d}^2n_s/{\rm d(ln}k)^2$, are different in the two theories. Therefore, if the SM Higgs indeed is the inflaton, detailed measurements of the inflationary observables can be used to distinguish between different gravitational degrees of freedom.

The above discussion applies only at classical level. Incorporating quantum corrections is a difficult task, although in the case of plateau potentials they have been shown to be mostly insignificant during inflation \cite{Bilandzic:2007nb,DeSimone:2008ei,Bezrukov:2009db,Lerner:2009xg,George:2013iia,George:2015nza,Bezrukov:2014ipa,Saltas:2015vsc,Bezrukov:2014bra,Herranen:2016xsy,Fumagalli:2016lls,Fumagalli:2016sof,Bezrukov:2017dyv,Markkanen:2017tun}. However, they may affect the potential in the regime where reheating occurs \cite{Bezrukov:2014bra,Bezrukov:2014ipa}, and certainly one should find a way to relate the couplings at the scale of inflation to the physics at lower energies, in particular around the EW scale. This is not totally unfounded, as recently studied in Refs. \cite{Rasanen:2017ivk,Shaposhnikov:2020geh} in the context of Palatini-Higgs inflation. For other recent studies on quantum corrections in Higgs inflation, see Refs. \cite{Hamada:2014iga,Cook:2014dga,Bezrukov:2014ipa,Fumagalli:2016lls,Enckell:2016xse,Bezrukov:2017dyv,Enckell:2018kkc,Rasanen:2017ivk}. 

Finally, we note that taking either the canonical metric or Palatini approach as a starting point is not the only way to modify this simplest kind of inflationary models. Other classes of Higgs-type inflation based on e.g. non-canonical kinetic couplings or teleparallel gravity have been considered in Refs. \cite{Germani:2010gm,Nakayama:2010kt,Kamada:2010qe,Kamada:2012se,Jinno:2017lun,Raatikainen:2019qey}. Also, we note that qualitatively nothing changes in the above discussion if the inflaton is not the SM Higgs but some other self-interacting scalar with a $\lambda\phi^4$ potential. Such scalar would necessarily belong to the BSM sector and could therefore be connected to other unresolved problems in modern physics, such as dark matter \cite{Lerner:2009xg,Lerner:2011ge,Bastero-Gil:2015lga,Kahlhoefer:2015jma,Tenkanen:2016twd,Heurtier:2017nwl,Hooper:2018buz,Almeida:2018oid} or the nature of the electroweak phase transition \cite{Tenkanen:2016idg}. However, in practice many details, such as reheating and its effect on the number of e-folds, and therefore also the predictions of such models for inflationary observables would be different from the case where the SM Higgs is the inflaton field and which therefore allow one to distinguish between different models \cite{Lerner:2009xg,Lerner:2011ge,Takahashi:2018brt}. Therefore, inflation provides a unique probe on such models at high energies.

\subsubsection{$R^2$ inflation}

Let us then consider another model where a different choice of the underlying gravitational degrees of freedom leads to interesting differences between the usual metric case and its alternatives. We consider the famous $R^2$ or Starobinsky model \cite{Starobinsky:1980te}, which is one of the oldest inflationary models. The scenario is specified by the following gravity sector:
\begin{equation}
\label{OmegaStarobinsky}
\Omega^2 = 1 + \alpha\frac{R^2}{M_{\rm P}^2}\,,
\end{equation}
where $\alpha$ is a dimensionless parameter, and no extra scalars are assumed to be present.

Let us start by discussing the metric case. By using the standard tricks (see e.g. Ref. \cite{Takahashi:2018brt}), one can show that for the above choice of $\Omega$ the Einstein frame potential becomes
\be
\label{UStarobinsky}
U(\chi) = \frac{M_{\rm P}^4}{8\alpha} \left(1-e^{-\sqrt{\frac{2}{3}}\frac{\chi}{M_{\rm P}}} \right)^2 ,
\ee
which closely resembles the potential in the metric Higgs case\footnote{This is not a coincidence, as we will discuss in Sec. \ref{xi-attractors}.}, Eq.~(\ref{chipotential1}), for $\chi\gg \sqrt{3/2}M_{\rm P}$ and with the identification $\alpha = \xi^2/2\lambda$. Here $\chi$ is a scalar field -- often called a "scalaron" -- which in the Jordan frame is hidden inside the $R^2$ term but which in the Einstein frame becomes dynamical. Assuming that the scalaron is the only source of inflationary fluctuations\footnote{In principle, the SM Higgs should not be forgotten, and one can ask what is its effect on the inflationary dynamics. It can be shown that the presence of the Higgs alongside the $R^2$ term leads to multifield inflation in metric gravity, as recently studied in Refs. \cite{Salvio:2015kka,Calmet:2016fsr,Wang:2017fuy,Ema:2017rqn,He:2018gyf,Ghilencea:2018rqg,Gundhi:2018wyz,Karam:2018mft,Enckell:2018uic,Canko:2019mud}.}, the spectral index and tensor-to-scalar ratio become
\begin{eqnarray}
\label{nsr_starobinsky}
n_s - 1&\simeq& - \frac{2}{N} - \frac{9}{2N^2}\,, \\
r &\simeq& \frac{12}{N^2}+\frac{18}{N^3}  \,,
\end{eqnarray}
which are the same as in the case of metric Higgs inflation to first order in the expansion in $1/N$, and therefore in agreement with the data. The correct amplitude for the curvature power spectrum is obtained for 
\be
\alpha\simeq \frac{N^2}{144\pi^2\mathcal{A}} \,,
\ee
which for the usual number of e-folds gives $\alpha \sim 10^9$. However, while the predictions for inflationary observables match to those of the metric Higgs inflation to first order in the expansion in $1/N$, the number of e-folds between the horizon exit of the pivot scale and the end of inflation is not the same in the two models\footnote{This is due to the fact that the scalaron couples to matter differently than the Higgs. For details, see Refs. \cite{Gorbunov:2010bn,Bezrukov:2011gp}.}, and therefore the exact predictions will not be the same either, even at the leading order. Therefore, detailed measurements of inflationary observables can distinguish between these two models.

Let us then consider the Palatini case. In this case, the choice (\ref{OmegaStarobinsky}) leads to a scenario where the scalaron is not dynamical and inflation cannot happen \cite{Meng:2003bk,Meng:2004yf}. However, assuming that in addition to the $R^2$ term the Jordan frame action contains at least one extra scalar field $\phi$ which is dynamical (such as the SM Higgs) and couples to gravity through a term of the form $G(\phi)g^{\mu\nu}R_{\mu\nu}(\Gamma)$, a Weyl transformation leads to \cite{Enckell:2018hmo,Antoniadis:2018ywb}
\begin{equation}
\label{SE_R^2}
S_E = \int {\rm d}^4x \sqrt{-g}\bigg[\frac12 M_{\rm P}^2 R- \frac12\partial^\mu\chi\partial_\mu\chi
+ \frac{\alpha}{2}\left(1+8\alpha\frac{\bar{U}}{M_{\rm P}^4}\right)\left(\partial^\mu\chi\partial_\mu\chi\right)^2 - U(\chi)\bigg] ,
\end{equation}
where the re-defined inflaton field $\chi$ is given by
\begin{equation}
\frac{{\rm d}\phi}{{\rm d}\chi} =\sqrt{\left(1+G(\phi)\right)\left(1+8\alpha\frac{\bar{U}}{M_{\rm P}^4}\right)} ,
\end{equation}
and the potential for the field is
\begin{equation}
U(\chi) \equiv \frac{\bar{U}(\chi)}{1+8\alpha\bar{U}(\chi)/M_{\rm P}^4}\,, \quad
\bar{U}(\chi) \equiv \frac{V(\phi(\chi))}{\left(1+G(\phi(\chi)) \right)^2} .
\end{equation}
As is evident from Eq. (\ref{SE_R^2}), in this case the scalar field kinetic term in Eq. (\ref{SE_R^2}) is manifestly non-canonical and the model does not correspond to a usual single-field canonical inflation scenario. However, by assuming slow-roll, one can show that the non-canonical part of the kinetic term in Eq. (\ref{SE_R^2}) is subdominant and can be neglected\footnote{One can still ask how quickly the slow-roll regime is reached if the non-canonical terms are important in the beginning of inflation. This was recently addressed in Ref. \cite{Tenkanen:2020cvw}.}. That being the case, the usual slow-roll parameters become
\begin{eqnarray}
\epsilon &\equiv& \frac12 M_{\rm P}^2\left(\frac{U_{,\chi}}{U}\right)^2 =\frac{1}{2} \left(\frac{U_{,\phi}}{U} \frac{{\rm d}\phi}{{\rm d}\chi} \right)^2 = \frac{U}{\bar{U}}\bar{\epsilon}\,,\\ 
\eta &\equiv& M_{\rm P}^2\frac{U_{,\chi,\chi}}{U} = \bar{\eta} - 24\alpha U \bar{\epsilon}\,,
\end{eqnarray}
where the subscripts of $U$ denote its derivatives with respect to $\chi$ and $\phi$ as specified after the comma and the overbars denote the slow-roll parameters in the case where the $R^2$ term is absent ($\alpha = 0$), when $U=\bar{U}$ and ${\rm d}\phi/{\rm d}\chi = \sqrt{1+G(\phi)}$. The amplitude of the curvature power spectrum and the spectral tilt are then given by
\begin{eqnarray}
24\pi^2M_{\rm P}^4\mathcal{A} &=& \frac{U}{\epsilon} =\frac{\bar{U}}{\bar{\epsilon}} \\ 
n_s - 1&=& 2\eta - 6\epsilon = 2\bar{\eta} - 6\bar{\epsilon}\,.
\end{eqnarray}
Therefore, even though the $R^2$ term can modify the inflaton potential significantly, the effect cancels out in $\mathcal{A}$ and $n_s$. As was shown in Ref. \cite{Enckell:2018hmo}, this is also true for the observables defined as higher order derivatives of the curvature power spectrum, such as the running or running of the running of the spectral tilt, because the curvature power spectrum remains the
same. However, that is not the case for the tensor power spectrum
\begin{equation}
\mathcal{P}_T = \frac{2}{3\pi^2}\frac{U}{M_{\rm P}^4} = \frac{2}{3\pi^2}\frac{\bar{U}/M_{\rm P}^4}{1+8\alpha\bar{U}/M_{\rm P}^4} \,,
\end{equation}
and, as a result, the tensor-to-scalar ratio becomes
\begin{equation}
\label{r_general}
r = 16\epsilon = \frac{\bar{r}}{1 + 8\alpha \bar{U}/M_{\rm P}^4} .
\end{equation}
This has interesting consequences; in particular, the result shows that regardless of how stringent the limit on $r$ becomes in the future, this class of models will always remain compatible with the data for large enough $\alpha$. The result also allows one to construct classes of simple models which have a very low scale of inflation, $V^{1/4}\propto \sqrt{H_{\rm inf}} \propto r^{1/4}$, as was recently studied in Ref. \cite{Tenkanen:2019wsd}. See also Refs. \cite{Antoniadis:2018yfq,Tenkanen:2019jiq,Tenkanen:2019wsd,Gialamas:2019nly} for other studies on Palatini inflation with an $R^2$ term.

\subsubsection{$\xi$-attractors}
\label{xi-attractors}

Finally, let us consider scenarios where the non-minimal coupling function and the Jordan frame potential are given by the following simple polynomials:
\begin{eqnarray}
\label{xi_attractor_models}
\Omega^2 &=& 1 + \xi \left(\frac{\phi}{M_{\rm P}}\right)^n\,, \\
V(\phi) &=& \lambda M_{\rm P}^{4-2n}\phi^{2n}\,,
\end{eqnarray}
where $n>0$ and $\lambda$ is a parameter that is fixed by the requirement that the curvature power spectrum has the measured amplitude. From the above choices, we see that the scenario is a generalization of the Higgs inflation model, Eq. (\ref{Higgs_inflation_model}). 

As is easy to show with the tricks discussed in the above sections, in the metric case the above choices of $\Omega$ and $V(\phi)$ correspond to 
\begin{equation}
\label{xi-attractor-potential}
U^{({\rm M})}(\chi) \simeq \frac{\lambda \, M_P^{4}}{\xi^2} \left(1-e^{-2\sqrt{\frac{\xi}{1+6\xi}} \frac{\chi}{M_P}} \right) ,
\end{equation}
where the superscript refers to the metric case and the result applies at large field values. This is the potential for the famous $\xi$-attractor models\footnote{For their relation to the famous "$\alpha$-attractor" models \cite{Ferrara:2013rsa,Kallosh:2013yoa} and terminology, see Ref. \cite{Galante:2014ifa}.} \cite{Kaiser:2013sna,Kallosh:2013tua}, where the name "attractor" refers to the fact that in the limit $\xi\to\infty$ the inflationary observables tend to
\begin{eqnarray}
n_s^{(M)} - 1&=& -\frac{2}{N}\,, \\ 
r^{(M)}&=& \frac{12}{N^2}\,,
\end{eqnarray}
where the expressions apply to first order in the expansion in $1/N$ and the superscripts refer to the metric case. It is noteworthy to point out how close these results are to those in the metric $R^2$ and Higgs inflation scenarios: at first order in $1/N$, the results for $n_s$ and $r$ are the same in all models, and therefore they are all in agreement with the current data. We can conclude that for the usual number of inflationary e-folds ($N=60$), the values $n_s \simeq 0.967$ and $r\simeq 0.0033$ are universal attractors in the limit of a strong non-minimal coupling, $\xi \sim 1$. 

However, as was shown in Ref. \cite{Jarv:2017azx}, this is no longer true for Palatini gravity. For general $n$ it is difficult to find a simple expression for the potential $U(\chi)$ but for e.g.\ $n=2$ one finds
\begin{equation}
U^{({\rm P})}(\chi) \simeq \frac{\lambda M_{\rm P}^4}{\xi^2} \left(1-8 e^{-\frac{2 \sqrt{\xi} \chi}{M_{\rm P}}} \right) \,,
\end{equation}
as in the case of Palatini-Higgs inflation, Eq. (\ref{chipotential1_P}), while $n=1$ yields
\begin{equation}
U^{({\rm P})}(\chi) \simeq \frac{\lambda M_{\rm P}^4}{\xi^2} \left(1- \frac{8 M_{\rm P}^2}{(2 M_{\rm P} + \xi \chi)^2}\right) \,.
\end{equation}
Analogous power laws can be found also for other values of $n$. For the inflationary observables one finds
\begin{eqnarray}
n_s^{(P)} - 1&\simeq& -\left(1-\frac{2}{n}\right)\frac{1}{N}\,, \\
r^{(P)} &\simeq& 0\,,
\end{eqnarray}
which again apply at the first order in the expansion in $1/N$ and in the limit $\xi\to\infty$. The results show that in this case the tensor-to-scalar ratio is not bounded from below, and therefore the attractor behavior found within the metric theory gravity is lost in the Palatini case. This is entirely due to the fact that in the Palatini case the kinetic term for the inflaton field (\ref{einsteinframe1}) is different from that in the metric case, which causes the dynamics during inflation to be different, and therefore also the observables to deviate from each other. For more details, see the discussion in Refs. \cite{Galante:2014ifa,Jarv:2017azx}.


\section{A goal for future missions}
\label{future_missions}

Finally, let us briefly discuss a specific goal for future missions which aim at detecting primordial gravitational waves. For other recent papers on the topic, see Refs. \cite{Takahashi:2018brt,Kallosh:2019eeu,Kallosh:2019hzo}.

As we have discussed in Sec. \ref{models}, the limit $r\sim 0.001$ for the primordial tensor-to-scalar ratio is special for two reasons: not only it is the value that one of the simplest inflationary models, $R^2$ (or Starobinsky) inflation, predicts and which is the limit which the $\xi$-attractor models (including the metric Higgs inflation) approach in the limit of a large non-minimal coupling, it is also the limiting value for $r$ in simple models of inflation with a non-minimal coupling which are based on the metric theory of gravity. In models which are based on the Palatini theory of gravity, the predicted value for $r$ can easily be as low as $r\sim 10^{-13}$, or even (much) lower than that when also other modifications to the gravity sector are allowed.

Therefore, going down to $r \leq 0.001$ can tell us a lot about the underlying gravitational degrees of freedom. If primordial gravitational waves are not observed above this limit, it implies -- in the context of the models studied in this paper -- that the (classical) metric theory of gravity is excluded, while almost all Palatini models would still be perfectly compatible with the data. Therefore, $r = 0.001$ should be an important target for future CMB B-mode polarization experiments such as BICEP3 \cite{Wu:2016hul}, LiteBIRD~\cite{Matsumura:2013aja}, CMB-S4 \cite{Abazajian:2016yjj}, the Simons Observatory \cite{Simons_Observatory}, and PICO \cite{Hanany:2019lle}, which indeed are soon pushing the limit on the tensor-to-scalar ratio down to $r=5\times 10^{-4}$, or aiming at detecting $r$ above this limit. As we have shown, in this way they may be able provide for a way to distinguish between different theories of gravity in the context of certain well-motivated classes of inflationary models.


\section{Summary}
\label{summary}

Currently we do not know what the underlying, high energy gravitational degrees of freedom are. However, by studying in detail the dynamics of different inflationary models and their predictions for observables, one may be able to distinguish not only between different models of inflation but also different theories of gravity. In this paper, we have elaborated this in the context of Palatini gravity, where in addition to the space-time metric also the connection is treated as a set of independent variables.

Here we have considered some well-motivated but simple and mostly canonical single-field models only. As discussed in Sec. \ref{future_missions}, going down to $r \leq 0.001$ can discriminate Palatini theories from the metric ones, at least in the context of the scenarios studied in this paper. Therefore, $r = 0.001$ is an important target for future missions aiming at detecting primordial gravitational waves.

It seems justified to say that at this moment in time, we are only in the beginning of understanding the connection between inflation and different theories of gravity. For example, besides understanding the fine details of the models considered in this paper, things that remain poorly or only partly understood in the context of Palatini models of inflation include e.g. multifield scenarios, details of (p)reheating in different scenarios, initial conditions for inflation, quantum corrections to different models, as well as scenarios with non-vanishing torsion or otherwise extended gravity sector and the former aspects in the presence of the latter, and so on. While important first steps have already been taken in this direction, it is clear that future studies hold great potential for revealing us a lot more about the connection between inflation and different theories of gravity and will -- hopefully -- help us to finally trace the high energy theory of gravity.


\begin{acknowledgements}
I thank F\'{e}lix-Louis Juli\'{e} and Ryan McManus for useful discussions and the Simons foundation for funding. 
\end{acknowledgements}

%
%

\bibliographystyle{spmpsci}      
\bibliography{Palatini_intro.bib}   

%
%

\end{document}